\documentclass[aps,prd,reprint,superscriptaddress,nofootinbib,longbibliography]{revtex4-1}
\usepackage{amsmath,amssymb, amsthm,amstext}
\usepackage{natbib}
\usepackage{graphicx}
\usepackage{color}
\usepackage{array, enumerate}
\usepackage{bm}
\usepackage{multirow}
\usepackage[breaklinks,colorlinks,citecolor=blue]{hyperref}
\usepackage{braket}
\usepackage{txfonts}
\usepackage{physics}
\usepackage[normalem]{ulem}
\usepackage{float}
\usepackage{subcaption}
\usepackage[dvipsnames]{xcolor}
\usepackage[singlelinecheck=false]{caption}

\makeatletter
\renewcommand{\p@subtable}{}
\makeatother

\def\nn{\nonumber}
\def\be{\begin{equation}}
\def\ee{\end{equation}}
\newcommand{\comment}[1]{}

\newcommand{\del}{\nabla}
\newcommand{\re}{\text{Re}}
\newcommand{\im}{\text{Im}}

\newcommand{\call}{\ensuremath{\mathcal{L}} }

\newcommand{\calr}{\ensuremath{\mathcal{R}} }

\newcommand{\neh}{\ensuremath{\triangle} }
\newcommand{\nehcs}{\ensuremath{\tilde{\triangle}} }

\begin{document}

\title{Exact Mass Multipoles of Distorted Schwarzschild Spacetimes}

\author{Brennen Black}
\email{brennenblack@sjtu.edu.cn}
\affiliation{Tsung-Dao Lee Institute, Shanghai Jiao-Tong University, Shanghai, 520 Shengrong Road, 201210, China}
\affiliation{School of Physics \& Astronomy, Shanghai Jiao-Tong University, Shanghai, 800 Dongchuan Road, 200240, China}

\begin{abstract}
    Using the isolated-horizon framework, we compute the geometric and mass multipole moments of the Weyl-distorted Schwarzschild horizon without linearizing the distortion. Starting from the exact integral representation we evaluate it in closed form for two cases, a pure quadrupole distortion and a dipole-quadrupole-octupole distortion. For a pure quadrupole we obtain a closed form for all $\ell$ in terms of confluent hypergeometric functions, while for a dipole-quadrupole-octupole distortion we derive a closed-form series solution in the odd-parity coefficient $\beta$ whose coefficients are exact in the quadrupole coefficient $\alpha_{2}$. We use these results to understand the errors incurred by leading-order approximations. Leading-order approximations in $\alpha_{2}$ misestimate every dimensionless mass multipole by a relative error of roughly $\alpha_{2}$, nearly independent of $\ell$, and for pure linearization most of the error lies in the linearized quadrupole moment rather than the discarded higher multipoles. A simple degree-counting rule determines which orders of distortion contribute to each multipole. As a consequence, the odd-parity component must be small compared with $\alpha_{2}^2$ before it can be neglected. Finally, truncating the series in $\beta$ at its first $\beta$-dependent order reproduces the $\ell \leq 6$ multipoles to about $10^{-3}$ for $|\beta|,|\alpha_{2}| \leq 1/8$.
\end{abstract}
\date{\today}

\maketitle

\section{Introduction}\label{sec:level1}
Black holes in astrophysical environments are never truly isolated systems. Companions, accretion disks, gas clouds and other sources create external tides that tidally distort the black hole \cite{Zhang1986,Geroch1982,Grlebeck2015}. The simplest systems are those that are static and axisymmetric. The Weyl-distorted Schwarzschild (WdS) spacetime is the class of spacetimes that include these symmetries \cite{Weyl1917,Geroch1982}. They are local vacuum solutions that describe the spacetime from the horizon to the external source within which the nonlinear Einstein equations reduce to a linear Laplace equation for a single function $U$ \cite{Weyl1917,Geroch1982,GriffithsPodolský2009}. The WdS spacetime is a standard sandbox for studying how external tides impact physical phenomena: companions \cite{Fang2005}, trapped surfaces \cite{Pilkington2011}, MOTS bifurcations \cite{Booth2026}, shadows \cite{Abdolrahimi2015a,Abdolrahimi2015b,Grover2018}, geodesics \cite{Shoom2016}, accretion disks \cite{Faraji2020}, the black hole interior \cite{Frolov2007}, and chaos \cite{Vieira1996}. In many cases the distortion associated with the external source is modeled as a pure quadrupole \cite{Shoom2016,Grover2018,Faraji2020,Booth2026}, however occasionally one considers up to octupole order \cite{Abdolrahimi2015a,Abdolrahimi2015b,Frolov2007,Vieira1996}.

Although the field multipole moments of a tidally deformed Schwarzschild black hole are unchanged by the tide, its horizon is not \cite{Grlebeck2015,Metidieri2024,Barcel2025}. Indeed it has been shown that the tidal Love numbers \cite{Love1909a,Love1909b,Flanagan2008,Hinderer2008} of 4-dimensional black holes (within the framework of General Relativity) vanish at linear order \cite{Fang2005,Binnington2009,Damour2009b,Kol2012,Cardoso2017,Hui2021,Charalambous2021}, beyond linear order \cite{Poisson2021,Riva2024,Iteanu2025} and, for static axisymmetric (Weyl) configurations, to all orders \cite{Kehagias2024,CombaluzierSzteinsznaider2025}. These results imply that the field multipoles of a distorted Schwarzschild black hole match those of an undistorted one. In contrast, the surficial Love numbers for the Weyl configuration have been shown to be non-vanishing, indicating that the horizon is tidally deformed \cite{Damour2009,Landry2014,Metidieri2024}. This deformation can similarly be characterized by a multipole decomposition of the horizon's curvature \cite{Ashtekar2004} and generally will not match the field multipoles \cite{Ashtekar2004,Gourgoulhon2026}.

Damour and Lecian \cite{Damour2009} derive an exact integral for the Legendre coefficient of the horizon curvature for an arbitrary Weyl distortion, but evaluated it only at linear order in the tidal field. The horizon multipoles of Refs.~\cite{Metidieri2024,Poisson2010} are likewise perturbative. The nonlinear multipoles have yet to be calculated, leaving a number of open questions unaddressed. Arguably the most relevant question is that of the error of linearization. Linearization necessarily discards information, and those wishing to use the linearized results for benchmarks in numerical studies \cite{Schnetter2006,Owen2009,PookKolb2020,Chen2022,Prasad2022,Prasad2024} need to address this. A second question is how a more general distortion is distributed among the horizon multipoles beyond linear order. For example, how small an odd-parity component of the external source must be before it can be neglected.

In this paper we aim to shed light on these questions by calculating in closed form the geometric ($I_{\ell}$) and mass ($M_{\ell}$) multipoles for a distortion of arbitrary strength. We do this by following the isolated-horizon framework developed by Ashtekar and collaborators \cite{Ashtekar2000,Ashtekar2002,Ashtekar2004,Ashtekar2025}. For the mass multipoles in particular they first define invariant multipoles $I_{\ell}$, obtained by projecting the real part of the Weyl scalar $\Psi_{2}$ onto the cross-sections of the horizon. On the horizon the relation $\re\Psi_{2} = -\calr/4$, with \calr the horizon curvature, shows that these geometric multipoles $I_{\ell}$ are related to the horizon multipoles by a factor of 4. Through simple dimensional analysis and normalization conditions, they then define the mass multipoles $M_{\ell}$. The invariance of $M_{\ell}$ follows from the invariance of $I_{\ell}$.

For a pure quadrupole distortion we obtain $I_{\ell}$ and $M_{\ell}$ in closed form for every $\ell$ and every distortion strength $\alpha_{2}$, while for a dipole-quadrupole-octupole distortion we obtain them as a series in the odd-parity coefficient $\beta$ whose coefficients are exact in $\alpha_{2}$. These results show that, while the field multipoles are unchanged by the tide, the horizon multipoles are non-zero at every even $\ell$, with $I_{\ell}$ first appearing at order $\alpha_{2}^{\ell/2}$. They also allow us to quantify the error of linearization directly. Leading-order approximations misestimate every dimensionless mass multipole of the pure quadrupole distortion by a relative error of roughly $\alpha_{2}$. For the dipole-quadrupole-octupole distortion a simple degree-counting rule determine which orders of the distribution contribute to each multipole.

The paper is outlined as follows. In Sec.~\ref{sec:level2} we provide a self-contained discussion of the WdS spacetime. We develop the distorted Schwarzschild metric from the canonical Weyl metric, discuss the functional form of the potential $U$, derive the conditions that are placed on the potential $U$ by metric regularity, and then calculate relevant horizon quantities. In Sec.~\ref{sec:level3} we review the exact integral solution for $I_{\ell}$ as obtained in the isolated-horizon framework. In Secs.~\ref{sec:level3.1} and \ref{sec:level3.2} we specialize the general formula to a pure quadrupole distortion and a dipole-quadrupole-octupole distortion, respectively. In Sec.~\ref{sec:level4} we use the exact results to investigate the relative errors incurred by taking leading-order approximations (of which linearization is a part) of of the multipoles, derive the rule that determines which orders of distortions contribute to each multipole, and quantify the error of truncating the series solution in $\beta$. Finally in Sec.~\ref{sec:level5} we summarize our work and provide discussion on the limitations and potential applications of this work. In this work we use units where $G = c = 1$ and take the metric signature to be $(-,+,+,+)$.

\section{The Weyl-distorted Schwarzschild Spacetime}\label{sec:level2}
\subsection{The Metric}\label{sec:level2.1}
In this section, we will review how to construct the Weyl metric. Let us begin by first imposing two symmetries: staticity and axisymmetry. With each of these symmetries comes a Killing vector field; one often denotes the static Killing vector by $\xi$ and the axisymmetric Killing vector by $\varphi$, and we will adopt these definitions. To each of these vectors we supply a coordinate adapted to the corresponding symmetry. For staticity we use the standard time coordinate $t\in(-\infty,\infty)$, and for axisymmetry we use the angular coordinate $\phi\in[0,2\pi)$. In such coordinates, one can express the Killing vectors as
\begin{equation}\label{eq:KillingVectorDefn}
    \xi = \partial_{t}\qquad\text{and}\qquad \varphi = \partial_{\phi}.
\end{equation}
For a complete chart of the 4-dimensional manifold let us introduce two more arbitrary coordinates $\eta$ and $\upsilon$. A general line element that is static and axisymmetric is then given by \cite{Geroch1982}
\begin{equation}\label{eq:GeneralWeylMetric}
    ds^2 = -e^{2\mathcal{U}}dt^2 + e^{-2\mathcal{U}}[e^{2\gamma}(d\eta^2 + d\upsilon^2) + \rho^2d\phi^2],
\end{equation}
where $\mathcal{U} = \mathcal{U}(\eta,\upsilon)$, $\gamma = \gamma(\eta,\upsilon)$ and $\rho = \rho(\eta,\upsilon)$. Let us now suppose there is a region of the spacetime which is a vacuum ($\Lambda = 0$). The field equations imply that $\rho(\eta,\upsilon)$ satisfies the two-dimensional Laplace equation $\rho_{,\eta\eta} + \rho_{,\upsilon\upsilon} = 0$, where we have introduced the notation that $\rho_{,\eta\eta} = \partial^2\rho/\partial\eta^2$ with an analogous definition for $\rho_{,\upsilon\upsilon}$. Being a solution to the Laplace equation makes $\rho$ a harmonic function, which implies there is a harmonic conjugate $z(\eta,\upsilon)$ related to $\rho$ by the Cauchy-Riemann equations
\begin{equation}\label{eq:HarmonicConjugateZ}
    z_{,\upsilon} = \rho_{,\eta}\qquad\text{and}\qquad z_{,\eta} = -\rho_{,\upsilon}.
\end{equation}
It is straightforward to show that $z(\eta,\upsilon)$ also satisfies the two-dimensional Laplace equation. It is convenient to then adopt $(\rho,z)$, known as Weyl's canonical coordinates, in place of $(\eta,\upsilon)$ to express the metric
\begin{equation}\label{eq:WeylMetric-CanonicalCoordinates}
    ds^2 = -e^{2\mathcal{U}}dt^2 + e^{-2\mathcal{U}}[e^{2\gamma}(d\rho^2 + dz^2) + \rho^2d\phi^2].
\end{equation}
One generally takes $\rho \in [0, \infty)$ and $z\in(-\infty,\infty)$, and we note that now $\mathcal{U} = \mathcal{U}(\rho,z)$ and $\gamma = \gamma(\rho,z)$ are the only remaining degrees of freedom.

Before we consider how to determine $\mathcal{U}$ and $\gamma$, let us consider a coordinate transformation and new definitions of $\mathcal{U}$ and $\gamma$
\begin{subequations}
    \begin{align}
        F &= 1 - \frac{2M}{r},\label{eq:F-Defn}\\
        \rho &= r\sqrt{1-\frac{2M}{r}}\sin\theta,\label{eq:Rho-CoordTrans}\\
        z &= (r-M)\cos\theta,\label{eq:z-CoordTrans}\\
        \mathcal{U} &= U + \frac{1}{2}\ln F,\label{eq:calU-CoordTrans}\\
        \gamma &= V - \frac{1}{2}\ln\left(\frac{(r-M)^2-M^2\cos^2\theta}{r^2F}\right),\label{eq:Gamma-CoordTrans}
    \end{align}
\end{subequations}
where $r\in[2M,\infty)$ and $\theta\in[0,\pi]$. When the above transformation is plugged into Eq.~\eqref{eq:WeylMetric-CanonicalCoordinates}, we obtain a new metric
\begin{equation}\label{eq:WdSMetric}
\begin{split}
    ds^2 = -e^{2U}Fdt^2 + e^{-2U+2V}\left[\frac{dr^2}{F} + r^2d\theta^2\right] \\
    + e^{-2U}r^2\sin^2\theta d\phi^2.
\end{split}
\end{equation}
It is now clear that this spacetime is a black hole spacetime, in particular a distorted Schwarzschild spacetime with a horizon at $r = 2M$, with $M$ being the Komar mass of the black hole. If both $U = U(r,\theta)$ and $V = V(r,\theta)$ vanish the spacetime becomes exactly Schwarzschild. The metric of Eq.~\eqref{eq:WdSMetric} will be the metric used for the remainder of the paper, and we will refer to this spacetime as the Weyl-distorted Schwarzschild (WdS) spacetime.

\subsection{The Weyl Potentials $U(r,\theta)$ and $V(r,\theta)$}\label{sec:level2.2}
The information regarding the distortion of the WdS spacetime from pure Schwarzschild are completely contained within the potentials $U$ and $V$. As it turns out and as we will now show, in a vacuum the only degree of freedom describing the distorted environment is within $U$. By solving the vacuum field equations, particularly $R_{tt} = 0$ and $R_{\phi\phi} = 0$, one finds that $U$ satisfies \cite{Geroch1982,GriffithsPodolský2009,Booth2026}
\begin{equation}\label{eq:WeylU-Equation}
    \frac{\partial}{\partial r}\left(r^2FU_{,r}\right) + \frac{1}{\sin\theta}\frac{\partial}{\partial\theta}(\sin\theta U_{,\theta}) = 0.
\end{equation}
The above equation can be regarded as the massless Klein-Gordon equation for a static $U(r,\theta)$ on the Schwarzschild spacetime, as it is equivalent to $r^2\Box U = 0$ on a Schwarzschild background (where $\Box := \del^{\mu}\del_{\mu}$ is the d'Alembertian operator). It yields an exact series solution in terms of Legendre polynomials \cite{Geroch1982,GriffithsPodolský2009,Booth2026}
\begin{equation}\label{eq:WeylU-Solution}
    U(r,\theta) = \sum_{n\geq1}\alpha_{n}\left(\frac{R}{M}\right)^{n}\mathcal{P}_{n}(r,\theta)
\end{equation}
where $\alpha := \{\alpha_{n}\}_{n\geq1}$ are the distortion coefficients, $R$ is defined by
\begin{equation}\label{eq:RDefn}
    R = \sqrt{r^2F + M^2\cos^2\theta},
\end{equation}
and the $\mathcal{P}_{n}$ are the Legendre polynomials of a modified argument
\begin{equation}\label{eq:PnDefn}
    \mathcal{P}_{n}(r,\theta) = P_{n}\left(\frac{(r-M)\cos\theta}{R}\right).
\end{equation}
While in general Eq.~\eqref{eq:WeylU-Solution} has a constant $n=0$ term, we omit it to fix the normalization of $\xi$. In many physical applications the distortion is modeled as a pure quadrupole ($\alpha_{2} \neq 0$ while all other coefficients vanish) \cite{Pilkington2011,Abdolrahimi2015a,Abdolrahimi2015b,Shoom2016,Faraji2020}, though a number of papers have gone as far as octupole distortions ($\alpha_{1},\alpha_{2},\alpha_{3} \neq 0$ while all other coefficients vanish) \cite{Frolov2007,Abdolrahimi2015a,Abdolrahimi2015b,Vieira1996}.

To obtain an equation for $V(r,\theta)$, one solves $R_{rr} = 0$, $R_{r\theta} = 0$ and $R_{\theta\theta} = 0$. One obtains the differential equations \cite{Geroch1982,GriffithsPodolský2009,Booth2026}
\begin{subequations}
    \begin{align}
        V_{,r} &= \frac{1}{\Xi}((r-M)A\sin\theta + B\cos\theta)\label{eq:WeylVr-Equation}\\
        V_{,\theta} &= \frac{1}{\Xi}((r-M)B\sin\theta - r^2FA\cos\theta)\label{eq:WeylVtheta-Equation}
    \end{align}
\end{subequations}
where
\begin{subequations}
    \begin{align}
        \Xi &= (r-M)^2 - M^2\cos^2\theta,\label{eq:Xi-Defn}\\
        A &= (r^2FU_{,r}^2 + 2MU_{,r} - U_{,\theta}^2)\sin\theta,\label{eq:A-Defn}\\
        B &= (2r^2FU_{,r} + 2M)U_{,\theta}\sin\theta.\label{eq:B-Defn}
    \end{align}
\end{subequations}
Eqs.~\eqref{eq:WeylVr-Equation} and \eqref{eq:WeylVtheta-Equation} are certainly formidable partial differential equations, but are not impossible to solve. While we will not provide the full calculation, we will describe the general steps. First, one recognizes that $V(r,\theta)$ is completely determined by $U(r,\theta)$, for which we already have a solution. Given that $V_{,r}$ and $V_{,\theta}$ have both linear and quadratic dependence on $U(r,\theta)$ (specifically its $r$- and $\theta$-derivatives), we can infer that $V(r,\theta)$ will have the following functional form
\begin{equation*}
    V(r,\theta) = V_{0}(r,\theta) + \sum_{i\geq1}\alpha_{i}V_{i}(r,\theta) + \sum_{j,k\geq1}\alpha_{j}\alpha_{k}V_{jk}(r,\theta).
\end{equation*}
By substituting this functional form back into the partial differential equations and matching powers of the distortion coefficients, one obtains partial differential equations for $V_{0}$, $V_{i}$ and $V_{jk}$ which can each be treated separately. One immediately finds that $V_{0}$ must be a constant $V_{0} = c_{0}$. $V_{i}(r,\theta)$ can be solved for by using recurrence relations and telescoping series, while $V_{jk}$ can be solved for by identifying that it is a homogeneous function of degree $j+k$ when expressed in Weyl's canonical coordinates. One is therefore able to determine that
\begin{align}\label{eq:WeylV-Solution}
    &V(r,\theta) = \tilde{c} - \frac{1}{M}\sum_{i\geq1}\alpha_{i}\bigg(\sum_{j=0}^{i-1}\bigg[(-1)^{i+j}(r-M(1-\cos\theta))\nn\\
    &\qquad\qquad\qquad\quad+ r - M(1+\cos\theta)\bigg]\left(\frac{R}{M}\right)^{j}\mathcal{P}_{j}\bigg)\nn\\
    &+ \sum_{j,k\geq1}\alpha_{j}\alpha_{k}\bigg(\frac{jk}{j+k}\left(\frac{R}{M}\right)^{j+k}\bigg[\mathcal{P}_{j}\mathcal{P}_{k} - \mathcal{P}_{j-1}\mathcal{P}_{k-1}\bigg]\bigg),
\end{align}
where the constant $\tilde{c}$ is defined for a fixed set of distortion coefficients via
\begin{equation*}
    \tilde{c} = c_{0} + \sum_{i\geq1}\alpha_{i}c_{i} + \sum_{j,k\geq1}\alpha_{j}\alpha_{k}c_{jk},
\end{equation*}
$c_{i}$ and $c_{jk}$ being the arbitrary integration constants for each $V_{i}$ and $V_{jk}$, respectively. This is not the final form of $V(r,\theta)$ as it contains an arbitrary constant $\tilde{c}$. In Sec.~\ref{sec:level2.3} we show that this constant must vanish.

In this work we will be particularly interested in the values that $U$ and $V$ take on the horizon (at $r = 2M$). We define these ''horizon quantities" with an overline, and they are given by
\begin{equation}\label{eq:WeylU-HorizonValue}
    \overline{U} := U(2M,\theta) = \sum_{n\geq1}\alpha_{n}\cos^{n}\theta,
\end{equation}
and
\begin{equation}\label{WeylV-HorionValue}
    \overline{V} := V(2M,\theta) = 2\overline{U} - 2\alpha_{\rm even}
\end{equation}
where we have defined $\alpha_{\rm even} = \sum_{i\geq1}\alpha_{2i}$ and anticipated the result $\tilde{c} = 0$ (Sec.~\ref{sec:level2.3}).

\subsection{Metric Regularity Conditions}\label{sec:level2.3}
Even though Eqs.~\eqref{eq:WdSMetric}, \eqref{eq:WeylU-Solution} and \eqref{eq:WeylV-Solution} solve the Einstein field equations, they are not guaranteed to give a regular solution. In particular, we have not yet enforced that the metric be smooth at the poles $\theta = 0,\pi$ of the symmetry axis. More rigorously, we must require that the manifold be locally flat at the poles, which means that the ratio of the circumference of closed curves around the symmetry axis to the proper distance from the symmetry axis should approach $2\pi$ as $\theta\to0,\pi$. If this ratio differs from $2\pi$ then the spacetime contains a conical singularity at the poles and so is not locally flat.

Let us consider a constant $(t,r)$ slice of the spacetime for some $r > 2M$. The induced metric is given by
\begin{equation*}
    d\omega^2 = r^2e^{-2U}(e^{2V}d\theta^2 + \sin^2\theta d\phi^2).
\end{equation*}
For a curve of constant $\theta$, the circumference is given by $C(\theta) = 2\pi re^{-U}\sin\theta$ while the proper distance is given by the solution $s(\theta)$ to $ds/d\theta = re^{-U+V}$. By requiring that the limit of $C(\theta)/s(\theta)$ goes to $2\pi$ as $\theta\to 0,\pi$ we obtain two conditions
\begin{align*}
    \lim_{\theta\to0,\pi}&\frac{C(\theta)}{s(\theta)} = \frac{\left|\frac{dC}{d\theta}\right|_{\theta=0,\pi}}{\left|\frac{ds}{d\theta}\right|_{\theta=0,\pi}} = 2\pi e^{-V(r,\theta=0,\pi)} = 2\pi\nn\\\nn\\
    &\implies \lim_{\theta\to0,\pi}V(r,\theta) = 0.
\end{align*}
These conditions constrain only $V$, not $U$. Evaluating Eq.~\eqref{eq:WeylV-Solution} at $\theta = 0,\pi$ gives the two conditions in algebraic form
\begin{align*}
    V(r,0) &= \tilde{c} + 2\sum_{i\geq1}\alpha_{2i-1} = 0,\quad \text{ and }\nn\\
    V(r,\pi) &= \tilde{c} - 2\sum_{i\geq1}\alpha_{2i-1} = 0
\end{align*}
which are only satisfied when
\begin{equation}\label{eq:MetricRegularityCondition}
    \tilde{c} = \sum_{i\geq1}\alpha_{2i-1} = 0.
\end{equation}
We see now that in addition to requiring $\tilde{c} = 0$, metric regularity forbids the existence of an isolated odd-parity distortion. Equivalently, $\alpha_{1}$ is not a free parameter; it is constrained by Eq.~\eqref{eq:MetricRegularityCondition}. We are therefore limited to the types of distortions we can analyze --- there must be at least two non-zero odd-parity distortion coefficients if one is to consider odd-parity distortions at all.

\subsection{The Horizon}\label{sec:level2.4}
The focus of this work is to understand the horizon of the WdS spacetime through recently developed, geometric-based formalisms which give the horizon multipoles and their Love numbers. In this section, we will introduce the preferred angular coordinate $\zeta$, which has been defined in \cite{Ashtekar2004}, as well as the intrinsic curvature of the horizon. In this section and for the remainder of the paper, the use of Greek letters (such as $\mu$ or $\nu$) as the indices of tensors will indicate that the tensor is defined on the full spacetime, lowercase Latin letters ($a$, $b$, $i$, $j$, $k$ and so on) will denote that the tensor is defined on a 3-dimensional submanifold of the spacetime, while uppercase Latin letters (such as $A$ or $B$) indicate that the tensor is defined on a 2-dimensional submanifold of the spacetime.

Let us first consider the null hypersurface \neh defined by $r = 2M$. The metric $q_{ab}$ of \neh induced by the spacetime metric $g_{\mu\nu}$ is given by
\begin{equation}\label{eq:HorizonMetric}
    q_{ab}dx^{a}dx^{b} = 4M^2e^{-2\overline{U}}\bigg[e^{2\overline{V}}d\theta^2 + \sin^2\theta d\phi^2\bigg],
\end{equation}
which is a degenerate metric (as is clear from the absence of a $dt^2$ term). The null hypersurface \neh is an axisymmetric Killing horizon and this comes with many important consequences, one of which is that for any null normal $\ell$ to \neh, $\call_{\ell}q_{ab} = 0$. This statement says that the horizon is non-expanding, which effectively means that any cross-section of the horizon is equivalent to any other cross-section. Let us therefore consider any cross-section \nehcs. The metric $\tilde{q}_{AB}$ of \nehcs induced by $q_{ab}$ takes the same form as Eq.~\eqref{eq:HorizonMetric}
\begin{equation}\label{eq:HorizonCrossSectionMetric}
    \tilde{q}_{AB}dx^{A}dx^{B} = 4M^2e^{-2\overline{U}}\bigg[e^{2\overline{V}}d\theta^2 + \sin^2\theta d\phi^2\bigg]
\end{equation}
but is not a degenerate metric; it is Riemannian. This metric supplies us with a unique derivative operator $\tilde{D}$ that is torsion-free and compatible with $\tilde{q}_{AB}$, along with an area 2-form $\tilde{\epsilon}$ that is Lie-dragged along the null normal $\ell$: $\call_{\ell}\tilde{\epsilon} = 0$ \cite{Ashtekar2000,Ashtekar2002,Ashtekar2004}. In principle there is also a rotation 1-form $\tilde{\omega}_{A}$ defined on \nehcs that is invariant under rescalings of the null normal $\ell$. For the WdS spacetime the rotation 1-form is exact, which implies that $\im\Psi_{2} = 0$ and all the $L_{\ell}$ vanish. The rotation 1-form is therefore not important to this discussion.

The Lie-dragging of the area 2-form is necessarily a consequence of the equivalence of horizon cross-sections, but this also allows us to define and calculate the areal radius $R_{\neh}$ which is itself independent of the choice of cross-section
\begin{equation}\label{eq:ArealRadius}
    R_{\neh}^2 := \frac{1}{4\pi}\oint_{\nehcs}\tilde{\epsilon} \implies R_{\neh} = 2Me^{-\alpha_{\rm even}}.
\end{equation}
The areal radius $R_{\neh}$ serves as a length scale for the intrinsic curvature, and can therefore be used to define a dimensionless intrinsic curvature (which will be of use in a later section).

The horizon \neh of the WdS spacetime is equipped with an axisymmetric Killing vector $\varphi$, and the area 2-form is Lie-dragged along the orbits generated by $\varphi$: $\call_{\varphi}\tilde{\epsilon} = 0$. This implies that $\tilde{\epsilon}_{BA}\varphi^{B}$ is a closed 1-form, and therefore that it can be written as the exterior derivative of a scalar $\zeta$ which is itself Lie-dragged along the orbits generated by $\varphi$. The function $\zeta$ is uniquely defined by
\begin{equation}\label{eq:InvariantCoordinateEquations}
    \tilde{D}_{A}\zeta = \frac{1}{R_{\neh}^2}\tilde{\epsilon}_{BA}\varphi^{B}\quad\text{and}\quad \oint_{\nehcs}\zeta\tilde{\epsilon} = 0,
\end{equation}
and is invariant under rescalings of the null normal to \neh. It is straightforward to see that $\zeta = \cos\theta$ satisfies Eq.~\eqref{eq:InvariantCoordinateEquations}. For a general spacetime, $\zeta = \cos\theta$ may not be an invariant scalar field, especially if the spacetime is not axisymmetric.

Now using $\zeta = \cos\theta$ and Eq.~\eqref{eq:ArealRadius}, we can express the horizon cross-section metric $\tilde{q}_{AB}$ as
\begin{equation}\label{eq:HorizonCrossSectionMetric-InvariantCoordinate}
    \tilde{q}_{AB}dx^{A}dx^{B} = R_{\neh}^2\left[\frac{e^{\overline{V}}}{1-\zeta^2}d\zeta^2 + e^{-\overline{V}}(1-\zeta^2)d\phi^2\right],
\end{equation}
from which we can identify the function $f(\zeta) := |\varphi|^2/R_{\neh}^2 = e^{-\overline{V}}(1-\zeta^2)$. A number of conditions were derived on $f(\zeta)$ and $f'(\zeta)$ which ensure the horizon is smooth everywhere, including the poles \cite{Ashtekar2002,Ashtekar2004}. In particular, one observes that $f(\pm1) = 0$ which indicates that $\varphi = \partial_{\phi}$ vanishes on the axis. Furthermore, we can calculate
\begin{equation}
    \frac{df}{d\zeta} = e^{-\overline{V}}\left(-2\zeta - (1-\zeta^2)\overline{V}_{,\zeta}\right)
\end{equation}
which gives $f'(\pm1) = \mp2$, hence the horizon cross-section is locally flat at the poles.

From Eq.~\eqref{eq:HorizonCrossSectionMetric-InvariantCoordinate} we can also calculate the intrinsic curvature $\calr$, which can be made dimensionless by multiplying by $R_{\neh}^2$
\begin{equation}\label{eq:IntrinsicCurvature}
    R_{\neh}^2\calr = -\frac{d^2f}{d\zeta^2} = -\frac{d^2}{d\zeta^2}\left[(1-\zeta^2)e^{-2(\overline{U}-\alpha_{\rm even})}\right],
\end{equation}
where now we have that
\begin{equation*}
    \overline{U} = \sum_{n\geq1}\alpha_{n}\zeta^{n}.
\end{equation*}
Equation \eqref{eq:IntrinsicCurvature} is the starting point for both the calculation of the mass multipole moments (through the on-horizon relation $\Psi_{2} = -\calr/4$) and the error analysis, and is consistent with Eq.~(24) of Ref.~\cite{Damour2009}.

\section{Horizon Multipole Moments}\label{sec:level3}
In the works of \cite{Ashtekar2004}, two sets of multipoles were introduced. The first are the geometric multipoles, denoted by $I_{\ell} + iL_{\ell}$, which are obtained by projecting the spin-weight-0 Weyl scalar $\Psi_{2}(r=2M)$ onto the horizon cross-section \nehcs using the spherical harmonics $Y_{\ell}^{m}(\zeta,\phi)$ associated to the invariant angular coordinate $\zeta$. In particular, $I_{\ell}$ is obtained by projecting the real part of $\Psi_{2}$ while $L_{\ell}$ is obtained by projecting the imaginary part. $L_{\ell}$ is related to the angular momentum of the black hole and by staticity must vanish in the WdS spacetime. The second set of multipoles are the mass and angular momentum multipoles, $M_{\ell}$ and $J_{\ell}$ respectively. In the WdS spacetime $J_{\ell} \equiv 0$ so we reserve the use of $J_{\ell}$ for other quantities. $I_{\ell}$ is related to the mass multipole moments $M_{\ell}$ by \cite{Ashtekar2004}
\begin{equation}\label{eq:MassMultipoleEquation}
    M_{\ell} = \sqrt{\frac{4\pi}{2\ell+1}}\frac{M_{\neh}R_{\neh}^{\ell}}{2\pi}I_{\ell},
\end{equation}
where $M_{\neh}$ is the mass obtained by the first law of black hole mechanics. In the WdS spacetime, one finds this to be $M_{\neh} = Me^{-\alpha_{\rm even}}$. Using that $R_{\neh} = 2M_{\neh}$ we can define a dimensionless mass multipole via
\begin{equation}\label{eq:DimensionlessMassMultipole}
    \frac{M_{\ell}}{M_{\neh}^{\ell+1}} = \sqrt{\frac{4\pi}{2\ell+1}}\frac{2^{\ell}}{2\pi}I_{\ell}.
\end{equation}

The geometric multipole moments $I_{\ell}$ are explicitly given by
\begin{equation}\label{eq:GeometricMultipoleDefn}
    I_{\ell} = -\oint_{\nehcs} \re\Psi_{2}Y_{\ell}^{0}(\zeta)dS,
\end{equation}
where $dS = R_{\neh}^2d\zeta d\phi$ is the horizon cross-section's physical area element. We note that in general one would want to keep all $m$ spherical harmonics, but in the case of an axisymmetric horizon only the $m = 0$ harmonic gives a non-vanishing result. The harmonic $Y_{\ell}^{0}(\zeta)$ can be expressed as a Legendre polynomial and $\Psi_{2}$ can be recovered from Eq.~\eqref{eq:IntrinsicCurvature},
\begin{align*}
    Y_{\ell}^{0} &= \sqrt{\frac{2\ell+1}{4\pi}}P_{\ell}(\zeta),\quad\text{and}\\
    R_{\neh}^2\re\Psi_{2} &= -\frac{R_{\neh}^2\calr}{4} = \frac{1}{4}\frac{d^2}{d\zeta^2}\bigg[(1-\zeta^2)e^{-2(\overline{U}-\alpha_{\rm even})}\bigg],
\end{align*}
which means we can express the $I_{\ell}$ as
\begin{equation*}
    I_{\ell} = \pi\sqrt{\frac{2\ell+1}{4\pi}}\int_{-1}^{1}\frac{d^2}{d\zeta^2}\left[-\frac{1}{2}(1-\zeta^2)e^{-2(\overline{U}-\alpha_{\rm even})}\right]P_{\ell}(\zeta)d\zeta.
\end{equation*}
Integrating by parts twice, one finds an exact form for $I_{\ell}$
\begin{align}\label{eq:GeometricMultipoleExactForm}
    I_{\ell} &= \pi\sqrt{\frac{2\ell+1}{4\pi}}\bigg\{1 + (-1)^{\ell} \nn\\
    &\qquad\quad- \frac{1}{2}\int_{-1}^{1}e^{-2(\overline{U} - \alpha_{\rm even})}(1-\zeta^2)P_{\ell}''(\zeta)d\zeta\bigg\},
\end{align}
where the double-prime on $P_{\ell}''(\zeta)$ stands for differentiation with respect to the invariant coordinate $\zeta$. In Eq.~\eqref{eq:GeometricMultipoleExactForm} one can identify that $(1-\zeta^2)P_{\ell}''(\zeta)$ is a solution to the generalized Legendre equation of order 2, indicating that it can be expressed as a Legendre function of degree $\ell$ and order 2: $(1-\zeta^2)P_{\ell}''(\zeta) = \mathbb{P}_{\ell}^{2}(\zeta)$. Eq.~\eqref{eq:GeometricMultipoleExactForm} is consistent with Eq.~(28) of Ref.~\cite{Damour2009}.

Without specifying $\overline{U}$ a closed-form solution (that is, one without an integral or an infinite summation) is not possible. However, Eq.~\eqref{eq:GeometricMultipoleExactForm} provides a very easy and efficient way to calculate multipoles for an arbitrary terms in $\overline{U}$. In many cases, the distortion is treated perturbatively with the quadrupole ($\alpha_{2}$) being the dominant term. It would therefore be insightful to determine an exact closed-form for quadrupole distortions (one that needs no integration) and to investigate how $M_{\ell}$ varies as a function of the distortion strength.

\subsection{Quadrupole Distortion}\label{sec:level3.1}
In astrophysical environments, the distortion produced by accretion disks or binary black holes are well approximated by quadrupole distortions. For such systems, we let $\alpha_{2}$ be non-zero while every other distortion coefficient vanishes. In this section, we evaluate in closed form what Ref.~\cite{Damour2009} calculates via the linear distortion approximation.

We start with Eq.~\eqref{eq:GeometricMultipoleExactForm} with the substitution $-(\overline{U} - \alpha_{\rm even}) = \alpha_{2}(1-\zeta^2)$. The main work comes from evaluating the integral of Eq.~\eqref{eq:GeometricMultipoleExactForm}, so we focus on just the integral. Let us give it a definition
\begin{equation}\label{eq:J_Ell-Defn}
    J_{\ell}(\alpha_{2}) := \int_{-1}^{1}e^{2\alpha_{2}(1-\zeta^2)}(1-\zeta^2)P_{\ell}''(\zeta)d\zeta
\end{equation}
so that
\begin{equation}\label{eq:I_Ell-Using-J_Ell}
    I_{\ell} = \pi\sqrt{\frac{2\ell+1}{4\pi}}\bigg\{1 + (-1)^{\ell} - \frac{1}{2}J_{\ell}\bigg\}.
\end{equation}
We observe that the integration interval of $J_{\ell}$ is symmetric and the factor multiplying $P_{\ell}''(\zeta)$ is even. The parity of the integral is therefore governed by the parity of $P_{\ell}''(\zeta)$, which follows from the parity of $\ell$. If $\ell$ is an odd number, $J_{\ell}$ and therefore $I_{\ell}$ vanish. A pure quadrupole distortion can thus only produce non-zero multipoles for even $\ell$. Furthermore, since $P_{\ell}''(\zeta)$ is a degree $\ell-2$ polynomial one must have that $J_{0} = J_{1} \equiv 0$. These vanishing integrals imply that $I_{0} = \sqrt{\pi}$ (as is required by the Gauss-Bonnet theorem) and that $I_{1} = 0$ (as is required by the fact that $\zeta$ was obtained by enforcing the vanishing of the area dipole).

Using the Legendre equation, we can rewrite $(1-\zeta^2)P_{\ell}''(\zeta) = 2\zeta P_{\ell}'(\zeta) - \ell(\ell+1)P_{\ell}(\zeta)$, which contains only one derivative term proportional to $P_{\ell}'(\zeta)$. For this derivative term, we integrate by parts which gives
\begin{align*}
    J_{\ell} &= 2\left(1 + (-1)^{\ell}\right) \nn\\
    &- \int_{-1}^{1}[(\ell^2 + \ell + 2) - 8\alpha_{2}\zeta^2]e^{2\alpha_{2}(1-\zeta^2)}P_{\ell}(\zeta)d\zeta.
\end{align*}
If we define a new integral function
\begin{equation*}
    \mathcal{J}_{\ell}(\alpha_{2}) = \int_{-1}^{1}e^{-2\alpha_{2}\zeta^2}P_{\ell}(\zeta)d\zeta
\end{equation*}
then we can express $J_{\ell}$ as
\begin{align*}
    J_{\ell}(\alpha_{2}) &= 2\left(1 + (-1)^{\ell}\right) - (\ell^2 + \ell + 2)e^{2\alpha_{2}}\mathcal{J}_{\ell}(\alpha_{2})\\
    &\qquad\qquad\qquad\quad- 4\alpha_{2}e^{2\alpha_{2}}\frac{d\mathcal{J}_{\ell}}{d\alpha_{2}},
\end{align*}
which means we only need to solve explicitly a fairly standard Gaussian integral.

To evaluate $\mathcal{J}_{\ell}(\alpha_{2})$ we start by rewriting $P_{\ell}(\zeta)$ using Rodrigues' formula, which gives
\begin{equation*}
    \mathcal{J}_{\ell}(\alpha_{2}) = \frac{1}{2^{\ell}\ell!}\int_{-1}^{1}e^{-2\alpha_{2}\zeta^2}\frac{d^{\ell}}{d\zeta^{\ell}}\bigg[(\zeta^2-1)^{\ell}\bigg]d\zeta.
\end{equation*}
We can integrate by parts $\ell$ times. Every boundary term will disappear, since fewer than $\ell$ derivatives act on $(\zeta^2-1)^{\ell}$ so a factor of $(\zeta^2-1)$ always survives. This shifts the $\ell$ derivatives to the exponential
\begin{equation*}
    \mathcal{J}_{\ell}(\alpha_{2}) = \frac{(-1)^{\ell}}{2^{\ell}\ell!}\int_{-1}^{1}\frac{d^{\ell}}{d\zeta^{\ell}}\bigg[e^{-2\alpha_{2}\zeta^2}\bigg](\zeta^2 - 1)^{\ell}d\zeta.
\end{equation*}
We then Taylor expand the exponential and apply the $\ell$ derivatives. Since only even $\ell$ contribute we write $\ell = 2L$ with $L$ a positive integer. The remaining integral over $\zeta\in[-1,1]$ can be expressed as twice the integral over $\zeta\in[0,1]$. With all these steps together one has
\begin{align*}
    \mathcal{J}_{2L}(\alpha_{2}) &= \frac{1}{4^{L}(2L)!}\sum_{n=L}^{\infty}\frac{(-2\alpha_{2})^{n}}{n!}\frac{(2n)!}{(2n-2L)!} \\
    &\qquad\times 2\int_{0}^{1}\zeta^{2(n-L)}(\zeta^2-1)^{2L}d\zeta.
\end{align*}
The integral in the above expression can be identified, through a series of appropriate substitutions, to be a Beta function integral. Such an integral can be expressed in terms of the Gamma function $\Gamma(x)$
\begin{align*}
    \mathcal{J}_{2L}(\alpha_{2}) &= \frac{1}{4^{L}(2L)!}\sum_{n=L}^{\infty}\frac{(-2\alpha_{2})^{n}(2n)!}{n!(2n-2L)!}\\
    &\qquad\times\frac{\Gamma(2L+1)\Gamma\left(n-L+\frac{1}{2}\right)}{\Gamma\left(n+L+\frac{3}{2}\right)}.
\end{align*}
Somewhat surprisingly, this summation simplifies to a single confluent hypergeometric function (CHF)
\begin{align*}
    \mathcal{J}_{2L}(\alpha_{2}) &= (-2\alpha_{2})^{L}\frac{\Gamma\left(L + \frac{1}{2}\right)}{\Gamma\left(2L + \frac{3}{2}\right)}{}_{1}F_{1}\left[L+\frac{1}{2};2L+\frac{3}{2};-2\alpha_{2}\right]\\
    &= (-2\alpha_{2})^{L}\frac{\Gamma\left(L + \frac{1}{2}\right)}{\Gamma\left(2L + \frac{3}{2}\right)}e^{-2\alpha_{2}}{}_{1}F_{1}\left[L+1;2L+\frac{3}{2};2\alpha_{2}\right],
\end{align*}
where the second equality is obtained by using Kummer's transformation. The summation can be carried out using basic symbolic mathematical software such as Mathematica. The final piece of the puzzle is the derivative of $\mathcal{J}_{2L}(\alpha_{2})$ with respect to $\alpha_{2}$, which takes a simple form with the use of derivative relations for CHFs
\begin{align*}
    &\frac{d\mathcal{J}_{2L}(\alpha_{2})}{d\alpha_{2}} = \left(\frac{L-2\alpha_{2}}{\alpha_{2}}\right)\mathcal{J}_{2L}(\alpha_{2})\\
    &+ (-2\alpha_{2})^{L}e^{-2\alpha_{2}}\frac{\Gamma\left(L + \frac{1}{2}\right)}{\Gamma\left(2L+\frac{3}{2}\right)}\frac{4L+4}{4L+3}{}_{1}F_{1}\left[L+2;2L+\frac{5}{2};2\alpha_{2}\right].
\end{align*}
We can finally put everything together to obtain $J_{\ell}$ (and thus $I_{\ell}$) as an exact function of $\alpha_{2}$
\begin{widetext}
    \begin{equation}\label{eq:J_Ell-ExactForm}
        J_{\ell}(\alpha_{2}) = \begin{cases}
            0,&\text{ if $\ell$ is odd}\\
            4 - 4(-2\alpha_{2})^{\frac{\ell}{2}}\frac{\Gamma\left(\frac{\ell+3}{2}\right)}{\Gamma\left(\ell + \frac{3}{2}\right)}\bigg[\frac{\ell}{2}{}_{1}F_{1}\left[\frac{\ell}{2}+1;\ell+\frac{3}{2};2\alpha_{2}\right] + {}_{1}F_{1}\left[\frac{\ell}{2};\ell+\frac{3}{2};2\alpha_{2}\right]\bigg],&\text{ if $\ell$ is even}.
        \end{cases}
    \end{equation}
\end{widetext}

\begin{figure}[htb]
    \centering
    \includegraphics[width=1.0\linewidth]{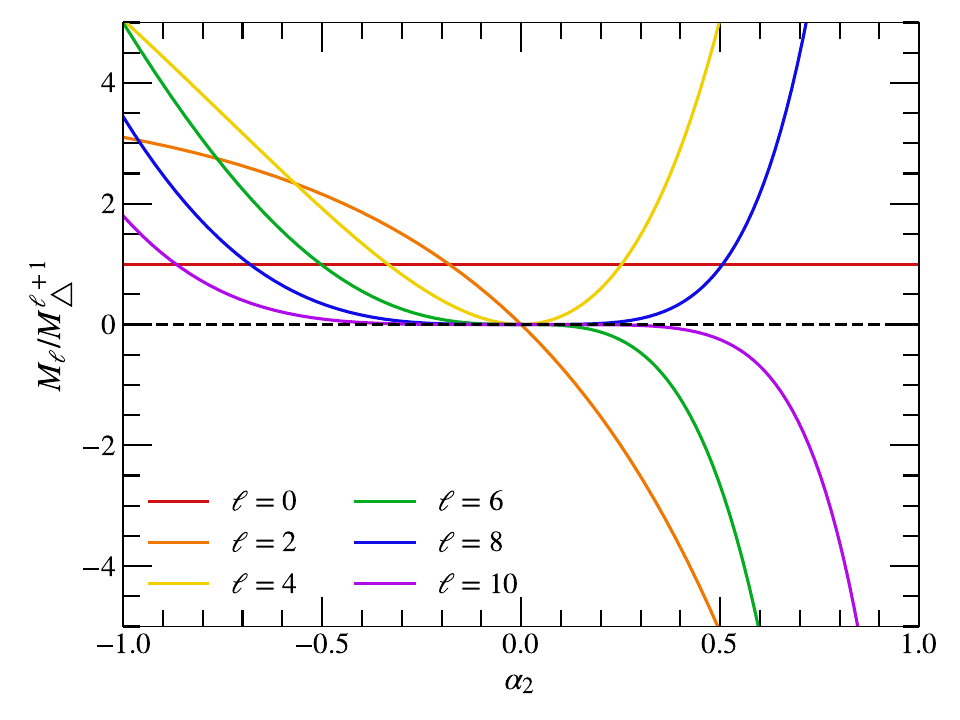}
    \caption{A plot of the dimensionless even-parity mass multipole moments for a pure quadrupole distortion from $\ell = 0$ to $\ell = 10$.}
    \label{fig:QuadrupoleMassMultipoleMoments}
\end{figure}

In Fig.~\ref{fig:QuadrupoleMassMultipoleMoments} we plot the dimensionless mass multipole moments $M_{\ell}/M_{\neh}^{\ell+1}$ for a pure quadrupole distortion. As is required by the Gauss-Bonnet theorem, $M_{0} = M_{\neh}$. When $\alpha_{2}\to+\infty$ each multipole moment except the monopole diverges. While it may not be obvious, when $\alpha_{2}\to-\infty$ each $M_{\ell}/M_{\neh}^{\ell+1}$ converges to $4^{\ell/2}$. This can be seen by considering the asymptotic behavior of Eq.~\eqref{eq:J_Ell-ExactForm}, in particular the behavior of the CHF ${}_{1}F_{1}[a;b;z]$ for large $|z|$ (which can be found in \cite{DLMF}).

Astrophysical black holes which are modeled using a pure quadrupole distortion generally have that $|\alpha_{2}|$ is smaller than unity. This is particularly true for those studies that take place outside the strong-field regime, in which case one can justify linearization in the distortion coefficient $\alpha_{2}$. From Eq.~\eqref{eq:J_Ell-ExactForm} it is clear that some error will be introduced by such an approximation. Furthermore, the leading order of Eq.~\eqref{eq:J_Ell-ExactForm} is $\alpha_{2}^{\ell/2}$, so linearization in $\alpha_{2}$ discards all $\ell > 2$ multipoles. In Sec.~\ref{sec:level4} we investigate how much information of the horizon curvature is lost or is erred by linearization in the distortion coefficient.

\subsection{Dipole-Quadrupole-Octupole Distortion}\label{sec:level3.2}
Beyond a pure-quadrupole distortion, some of the literature considers octupole-order distortions for analyzing different physical phenomena. A purely odd distortion is a rather unlikely picture as it requires a high degree of fine-tuning of the source generating the Weyl potential. In most astrophysical cases it is unlikely that the external source will be distributed in a way that is totally antisymmetric. A more likely scenario is an odd-parity perturbation atop a pure quadrupole. In this section we will derive a series solution for a dipole-quadrupole-octupole distortion of arbitrary strength.

We take the horizon potential to be
\begin{equation*}
    \overline{U} = \alpha_{2}\zeta^2 + \beta\zeta(1-\zeta^2),
\end{equation*}
where $\alpha_{2} = \alpha_{\rm even}$ is the quadrupole distortion coefficient and $\beta$ the odd-parity distortion coefficient. This amounts to taking $\alpha_{1} = -\alpha_{3} = \beta$, as required by Eq.~\eqref{eq:MetricRegularityCondition}. Much like the pure quadrupole case, the work comes in evaluating the following integral
\begin{equation}\label{eq:KEll-Defn}
    K_{\ell} = \int_{-1}^{1}e^{2(1-\zeta^2)(\alpha_{2}-\beta\zeta)}(1-\zeta^2)P_{\ell}''(\zeta)d\zeta
\end{equation}
which takes the place of $J_{\ell}$ from the previous section. The first step is to write a series expansion for $P_{\ell}''(\zeta)$. The Legendre polynomial has many series representations, but we find the easiest to use is
\begin{equation*}
    P_{\ell}(\zeta) = 2^{\ell}\sum_{k=0}^{\ell}\binom{\ell}{k}\binom{\frac{\ell+k-1}{2}}{\ell}\zeta^{k},
\end{equation*}
which, after taking two derivatives, gives a series expansion for $P_{\ell}''(\zeta)$
\begin{equation}\label{eq:PEllD2-SeriesExpansion}
    P_{\ell}''(\zeta) = 2^{\ell}\sum_{k=0}^{\ell-2}\binom{\ell}{k+2}\binom{\frac{\ell+k+1}{2}}{\ell}(k+1)(k+2)\zeta^{k}.
\end{equation}
Furthermore, we can split the exponential in the integrand of Eq.~\eqref{eq:KEll-Defn} into even and odd exponents (with respect to $\zeta$). The even exponential is a Gaussian exponential, while the odd exponential can be Taylor expanded
\begin{align}\label{eq:ExponentialSeries}
    e^{2(1-\zeta^2)(\alpha_{2}-\beta\zeta)} &= e^{2\alpha_{2}(1-\zeta^2)}e^{-2\beta\zeta(1-\zeta^2)}\nn\\
    &= e^{2\alpha_{2}(1-\zeta^2)}\sum_{n\geq0}\frac{(-2\beta)^{n}}{\Gamma(n+1)}\zeta^{n}(1-\zeta^2)^{n}.
\end{align}
Plugging Eqs.~\eqref{eq:PEllD2-SeriesExpansion} and \eqref{eq:ExponentialSeries} into Eq.~\eqref{eq:KEll-Defn} gives
\begin{widetext}
    \begin{equation*}
        K_{\ell} = \sum_{n\geq0}\sum_{k=0}^{\ell-2}\frac{(-2\beta)^{n}2^{\ell}}{\Gamma(n+1)}\binom{\ell}{k+2}\binom{\frac{\ell+k+1}{2}}{\ell}(k+1)(k+2)\int_{-1}^{1}e^{2\alpha_{2}(1-\zeta^2)}\zeta^{n+k}(1-\zeta^2)^{n+1}d\zeta.
    \end{equation*}
\end{widetext}
We observe that the integration interval is symmetric and the parity of the integrand follows from the parity of $n+k$. Instead of rewriting the summation indices to be even we apply the filter $(1+(-1)^{n+k})/2$ to each term, after which we are free to treat the integral as purely even. The remaining integral in the above equation can be shown, through a number of substitutions of the integration variable, to be a CHF
\begin{align}
    &\int_{-1}^{1}e^{2\alpha_{2}(1-\zeta^2)}\zeta^{n+k}(1-\zeta^2)^{n+1}d\zeta = \nn\\
    &\quad\quad\frac{\Gamma(n+2)\Gamma\left(\frac{n+k+1}{2}\right)}{\Gamma\left(\frac{3n+k+5}{2}\right)}{}_{1}F_{1}\left[n+2;\frac{3n+k+5}{2};2\alpha_{2}\right]
\end{align}
for $n+k$ even. Finally, putting these results together we obtain the series solution we sought
\begin{align}\label{eq:KEll-SeriesSolution}
    K_{\ell} &= 2^{\ell}\sum_{n\geq0}(-2\beta)^{n}(n+1)\nn\\
    &\times\sum_{k=0}^{\ell-2}\frac{1+(-1)^{n+k}}{2}\binom{\ell}{k+2}\binom{\frac{\ell+k+1}{2}}{\ell}(k+1)(k+2)\nn\\
    &\times\frac{\Gamma\left(\frac{n+k+1}{2}\right)}{\Gamma\left(\frac{3n+k+5}{2}\right)}{}_{1}F_{1}\left[n+2;\frac{3n+k+5}{2};2\alpha_{2}\right].
\end{align}
We note that when $\ell = 0,1$ the $k$-summation vanishes identically so $K_{0} = K_{1} = 0$, from which one deduces that $M_{0} = M_{\neh}$ and $M_{1} = 0$ as required. Figures \ref{fig:DPO_EvenLMoments} and \ref{fig:DPO_OddLMoments} show the first three even-$\ell$ and odd-$\ell$ mass multipoles beyond $\ell \geq 2$, respectively, for $-1/8\leq\beta\leq1/8$ and $\alpha_{2} \in \{-1/8, 0, 1/8\}$. A more detailed discussion of the elements of these plots is given in Sec.~\ref{sec:level4}.

\section{Accuracy of Leading-Order Multipoles}\label{sec:level4}
The existing literature evaluates the horizon multipoles in closed form only at linear order in the tidal field \cite{Damour2009} by treating $\alpha_{2}$ as both the dominant source of the distortion and as satisfying the condition $|\alpha_{2}| \ll 1$. Error is necessarily incurred by this approximation, but how much has not yet been quantified. In this section we consider a leading-order approximation, a more general form of pure linearization. As will be discussed, not every multipole has a linear term but that does not mean they vanish. Pure linearization therefore contributes two sources of error to the reconstruction of the horizon geometry, one from the approximation on $I_{2}$ (or equivalently $M_{2}$) and one from neglecting higher order multipoles.

We begin by examining the pure quadrupole distortion at leading order in $\alpha_{2}$. The leading term in the series expansion for the CHF is unity \cite{DLMF}, so the leading order of $J_{\ell}(\alpha_{2})$ (Eq.~\eqref{eq:J_Ell-ExactForm}) is
\begin{equation}\label{eq:LeadingOrder-J_Ell}
    J_{\ell}(\alpha_{2}) = 4 - 4(-2\alpha_{2})^{\ell/2}\left(1+\frac{\ell}{2}\right)\frac{\Gamma\left(\frac{\ell+3}{2}\right)}{\Gamma\left(\ell+\frac{3}{2}\right)} + \mathcal{O}\left(\alpha_{2}^{\ell/2 + 1}\right)
\end{equation}
where we have taken $\ell$ to be even. Using Eqs.~\eqref{eq:DimensionlessMassMultipole}, \eqref{eq:I_Ell-Using-J_Ell} and \eqref{eq:LeadingOrder-J_Ell} we determine the dimensionless mass multipoles to leading order in $\alpha_{2}$ for any even $\ell$ to be
\begin{equation}\label{eq:QuadDistortion-LeadingOrderMassMultipole}
    \frac{M_{\ell}^{\rm LO}}{M_{\neh}^{\ell+1}} = 2^{\ell}\left(1+\frac{\ell}{2}\right)\frac{\Gamma\left(\frac{\ell+3}{2}\right)}{\Gamma\left(\ell+\frac{3}{2}\right)}(-2\alpha_{2})^{\ell/2}
\end{equation}
while for odd $\ell$, $M_{\ell} \equiv 0$. As mentioned in Sec.~\ref{sec:level3.1}, linearization in $\alpha_{2}$ means each mass multipole for $\ell > 2$ vanishes since the leading-order behavior is $\alpha_{2}^{\ell/2}$. In general, a $k$-order expansion with respect to $\alpha_{2}$ will give $k$ non-zero mass multipoles beyond the $\ell = 0$ monopole.

Expanding the CHFs to the next order, ${}_{1}F_{1}[a;b;z] = 1 + az/b + \mathcal{O}(z^2)$ \cite{DLMF}, we find
\begin{align}\label{eq:QuadrupoleRelativeError}
    \frac{M_{\ell}}{M_{\neh}^{\ell+1}} &= \frac{M_{\ell}^{\rm LO}}{M_{\neh}^{\ell+1}}\left[1 + d_{\ell}\,\alpha_{2} + \mathcal{O}(\alpha_{2}^{2})\right],\nn\\
    &\text{where } d_{\ell} = \frac{2\ell(\ell+4)}{(\ell+2)(2\ell+3)} .
\end{align}
The relative error of the leading-order approximation to the next-order term (hereafter referred to as the relative error), $(M_{\ell}-M_{\ell}^{\rm LO})/M_{\ell}^{\rm LO}\simeq d_{\ell}\,\alpha_{2}$, is therefore nearly independent of $\ell$. The coefficient $d_{\ell}$ starts at $d_{2} = 6/7$ at $\ell = 2$, approaches unity as $\ell\to\infty$ and is never larger than 1.02 for $\ell \geq 2$. The leading order dimensionless mass multipole moments therefore misestimate the exact mass multipoles by roughly $\alpha_{2}$. For reference, at $\alpha_{2} = \pm0.1$ the relative errors in the $\ell = 2$ multipole are 9.1\% ($\alpha_{2} > 0$) and -8.1\% ($\alpha_{2} < 0$), while for $\ell = 16$ the relative errors are 10.7\% ($\alpha_{2} > 0$) and -9.6\% ($\alpha_{2} < 0$).

For pure linearization, these results separate the two sources of error identified at the start of this section. By Eq.~\eqref{eq:GeometricMultipoleDefn}, the $I_{\ell}$ are proportional to the components of the dimensionless curvature $R_{\neh}^2\calr$ along the orthonormal harmonics $Y_{\ell}^{0}$, so errors in different multipoles can be compared directly. At first order in $\alpha_{2}$, pure linearization misestimates $I_{2}$ by $d_{2}|\alpha_{2}||I_{2}| = (6/7)|\alpha_{2}||I_{2}|$, whereas the largest multipole it neglects is $|I_{4}| = (2\sqrt{5}/7)|\alpha_{2}||I_{2}|$, which follows from Eqs.~\eqref{eq:DimensionlessMassMultipole} and \eqref{eq:QuadDistortion-LeadingOrderMassMultipole}. Most of the error of pure linearization therefore lies in the one multipole it keeps rather than the multipoles it discards, and using the exact result of Eq.~\eqref{eq:J_Ell-ExactForm} removes this part of the error entirely.

These relative errors are not unique, and generally only follow from how one measures the tides. In fact they are only unambiguous at linear order, but there exists a set of justifiably canonical choices. Both the size of $d_{\ell}$ and its near independence of $\ell$ depend on two choices, the definition of the tidal amplitudes and the normalization of the mass multipoles. We have chosen to use the Weyl distortion coefficients $\alpha_{2}$ and $\beta$, defined with respect to the Komar mass $M$, as the expansion parameters. This choice is natural for two reasons. First, it is the choice under which the problem is linear, since $U$ is a linear superposition of the $\alpha_{n}$, both horizon metric functions $\overline{U}$ and $\overline{V}$ are linear, and so is the regularity condition. Second, these parameters are fixed by the horizon geometry alone. They can be obtained by reading the coefficients of the expansion of the horizon invariant $\overline{U}$ with respect to the invariant coordinate $\zeta$. For the pure quadrupole and dipole-quadrupole-octupole distortions specifically, $\alpha_{2}$ can equivalently be calculated using $\alpha_{2} = \ln(M/M_{\neh})$. We normalize by $M_{\neh}$ however so that the multipoles are expressed in units intrinsic to the horizon. Under these choices, the relative error is then a measure of the nonlinearity of the map from $\overline{U}$ to the horizon geometry, i.e. the exponential in Eq.~\eqref{eq:IntrinsicCurvature}.

Let us now consider the dipole-quadrupole-octupole distortion as calculated in Sec.~\ref{sec:level3.2}. Whereas reflection-symmetric sources lead to identically vanishing odd mass multipoles, one clearly sees from Eq.~\eqref{eq:KEll-SeriesSolution} that including odd-parity terms in the Weyl potential produces a full set of non-trivial odd mass multipoles. Such moments can be calculated by direct numerical integration of Eq.~\eqref{eq:GeometricMultipoleExactForm}, but the series of Eq.~\eqref{eq:KEll-SeriesSolution} gives closed-form coefficients of $M_{\ell}/M_{\neh}^{\ell+1}$ at each order in $\beta$ and with full $\alpha_2$ dependence, which is what allows us to quantify the error incurred by truncating in $\beta$.

To include a non-zero $\beta$ a natural first step is linearization in $\beta$, and a closed-form solution is readily achievable from Eq.~\eqref{eq:KEll-SeriesSolution} by truncating the $n$-summation appropriately. For a given $\ell$ multipole only those terms with $n\equiv\ell$ (mod 2) contribute, so the even $\ell$ modes only contain even powers of $\beta$ while the odd $\ell$ modes only contain odd powers of $\beta$. At linear order in $\beta$, the even $\ell$ mass multipoles are therefore exactly the same as those calculated in Sec.~\ref{sec:level3.1}; the odd $\ell$ multipoles are given by
\begin{align}\label{eq:BetaLinearOddMassMultipoles}
    \frac{M_{\ell}^{\rm(odd)}}{M_{\neh}^{\ell+1}} &= 2^{\ell-1}(\ell^2-1)\frac{\Gamma\left(\frac{\ell+4}{2}\right)}{\Gamma\left(\ell + \frac{3}{2}\right)}\beta (-2\alpha_{2})^{(\ell-3)/2}\nn\\
    &\times{}_{1}F_{1}\left[\frac{\ell+3}{2};\ell+\frac{3}{2};2\alpha_{2}\right] + \mathcal{O}(\beta^3).
\end{align}
This expression is valid for odd $\ell \geq 3$, but at $\ell = 1$ yields an indeterminate form. This is no pathology though, as $M_{1} \equiv 0$ for any distortion by construction. To match the leading-order discussion we gave for the pure quadrupole distortion, we will extend another degree in $\beta$ and present the closed form for the $\beta^2$ term. Let $M_{\ell}^{\rm(Q)}$ denote the exact solution for a pure quadrupole (equivalently the $\beta^{0}$ term), then the even multipoles are given by
\begin{align}\label{eq:QuadBetaEvenMassMultipole}
    &\frac{M_{\ell}^{\rm(even)} - M_{\ell}^{\rm(Q)}}{M_{\neh}^{\ell+1}} = 2^{\ell-3}\,\ell(\ell+2)\frac{\Gamma\left(\frac{\ell+3}{2}\right)}{\Gamma\left(\ell+\frac{3}{2}\right)}\beta^{2}(-2\alpha_{2})^{\frac{\ell}{2}-3}\nn\\
    &\qquad\qquad \times \bigg[\left(\ell^{2}+\ell+1-4\alpha_{2}\right){}_{1}F_{1}\!\left(\tfrac{\ell}{2}+2;\ell+\tfrac{3}{2};2\alpha_{2}\right)\nn\\
    &\qquad\qquad- 7(\ell-1)\,{}_{1}F_{1}\!\left(\tfrac{\ell}{2}+1;\ell+\tfrac{3}{2};2\alpha_{2}\right)\bigg] + \mathcal{O}(\beta^4).
\end{align}
The above equation has no problems for even $\ell \geq 6$, however for $\ell = 2,4$ it becomes an indeterminate form at $\alpha_{2} = 0$. One can use the equivalent forms
\begin{align}\label{eq:QuadBetaL2}
    \frac{M_{2}^{\rm(even)}-M_{2}^{\rm(Q)}}{M_{\neh}^{3}} &= \frac{64}{105}\beta^2\bigg[8{}_{1}F_{1}\left[5;\frac{11}{2};2\alpha_{2}\right]\nn\\
    &\quad- 9{}_{1}F_{1}\left[4;\frac{9}{2};2\alpha_{2}\right]\bigg] + \mathcal{O}(\beta^4)
\end{align}
and
\begin{align}\label{eq:QuadBetaL4}
    \frac{M_{4}^{\rm(even)}-M_{4}^{\rm(Q)}}{M_{\neh}^{5}} &= \frac{256}{231}\beta^{2}\bigg[28{}_{1}F_{1}\left[5;\frac{13}{2};2\alpha_{2}\right]\nn\\
    &\quad- 33{}_{1}F_{1}\left[4;\frac{11}{2};2\alpha_{2}\right]\bigg] + \mathcal{O}(\beta^4)
\end{align}
for these multipoles.

\begin{figure*}[tb]
    \centering
    \includegraphics[width=1.0\linewidth]{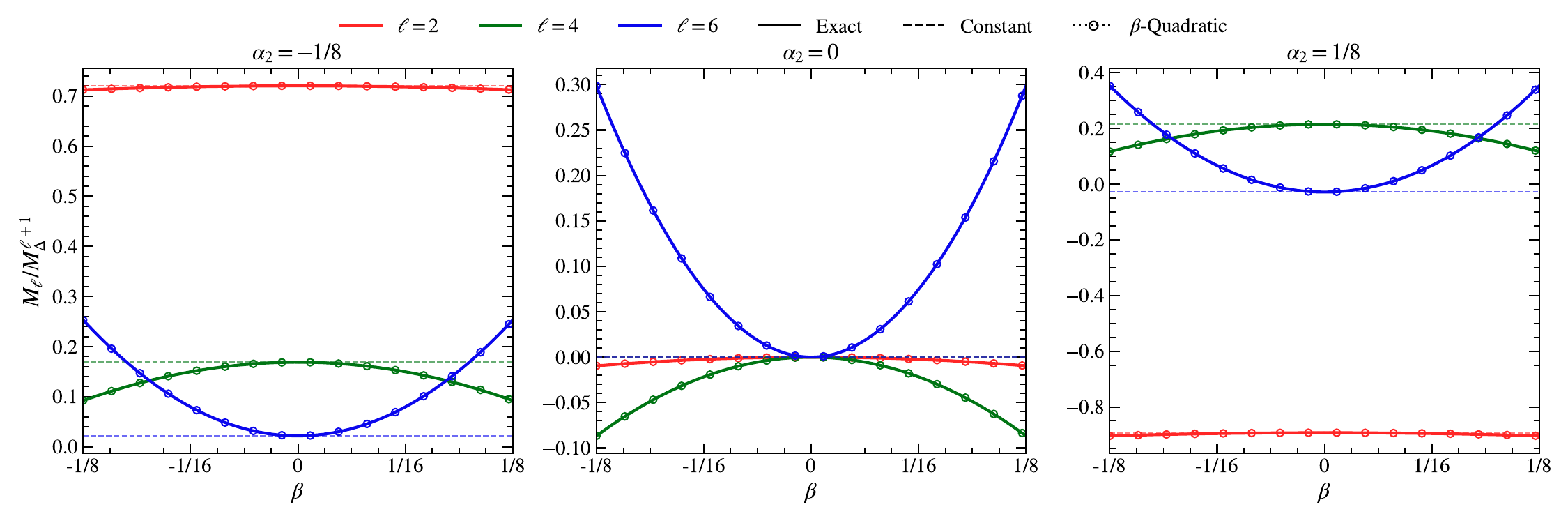}
    \caption{A plot of the dimensionless mass multipoles for the $\ell = 2$ (red), $\ell = 4$ (green) and $\ell = 6$ (blue) multipoles as a function of $\beta$ for a fixed $\alpha_{2}$ (left column $\alpha_{2} = -1/8$, center column $\alpha_{2} = 0$, right column $\alpha_{2} = 1/8$).}
    \label{fig:DPO_EvenLMoments}
\end{figure*}

\begin{figure*}[tb]
    \centering
    \includegraphics[width=1.0\linewidth]{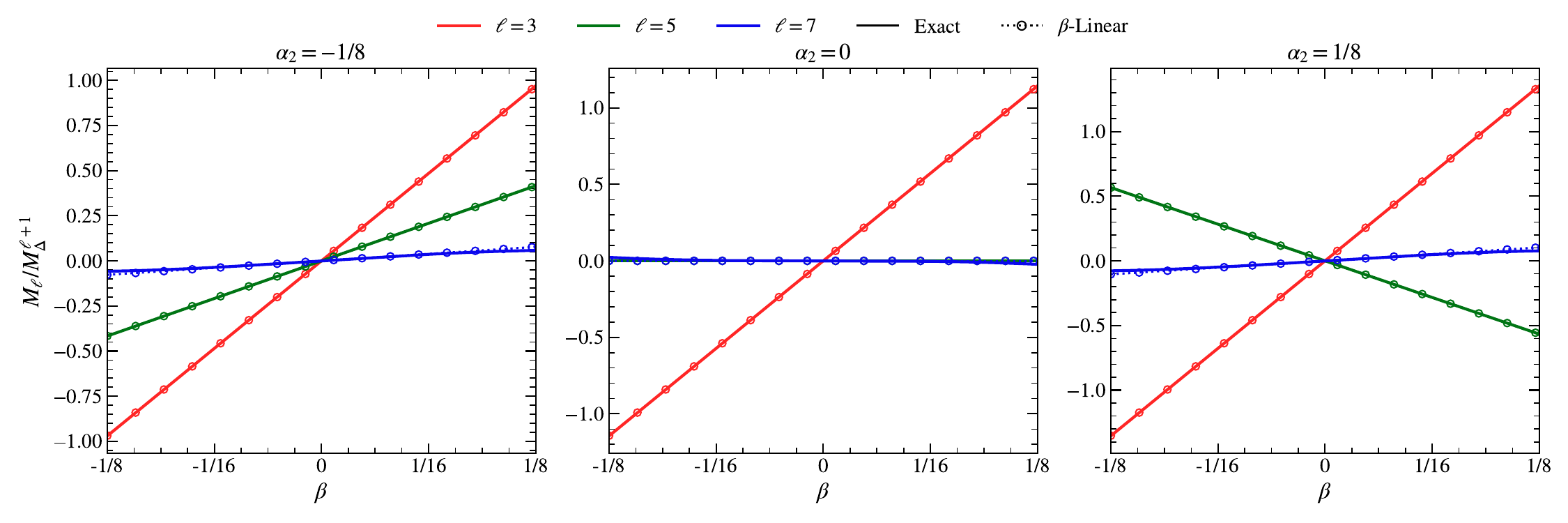}
    \caption{A plot of the dimensionless mass multipoles for the $\ell = 3$ (red), $\ell = 5$ (green) and $\ell = 7$ (blue) multipoles as a function of $\beta$ for a fixed $\alpha_{2}$ (left column $\alpha_{2} = -1/8$, center column $\alpha_{2} = 0$, right column $\alpha_{2} = 1/8$). At $\alpha_{2} = 0$, one has that $M_{5} \equiv 0$. This is due to the fact that the integrand of Eq.~\eqref{eq:KEll-Defn} can be expressed as a total derivative of a function which takes the same value at $\zeta = \pm1$.}
    \label{fig:DPO_OddLMoments}
\end{figure*}

Figures \ref{fig:DPO_EvenLMoments} and \ref{fig:DPO_OddLMoments} show the mass multipole moments for the dipole-quadrupole-octupole distortion for $\ell=2,4,6$ and $\ell=3,5,7$ (respectively) for $\alpha_{2}\in\{-1/8,0,1/8\}$ and $|\beta| \leq 1/8$. These figures compare the exact mass multipoles, obtained by numerical integration, to those multipoles obtained by the truncation of Eq.~\eqref{eq:KEll-SeriesSolution} up to $n = 2$, namely Eq.~\eqref{eq:BetaLinearOddMassMultipoles} for the odd-$\ell$ mass multipoles and Eqs.~\eqref{eq:QuadBetaEvenMassMultipole}-\eqref{eq:QuadBetaL4} for the even-$\ell$ mass multipoles. The dashed curves in Fig.~\ref{fig:DPO_EvenLMoments} show the $\beta$-independent term alone, i.e. the pure quadrupole results of Sec.~\ref{sec:level3.1}. Figure \ref{fig:DPO_OddLMoments} shows that for small $\beta$ the odd-$\ell$ multipoles are well approximated by linearization in $\beta$, and that this holds for any small $\alpha_{2}$. As for the even-$\ell$ mass multipoles, Fig.~\ref{fig:DPO_EvenLMoments} shows that in the range $|\beta|,|\alpha_{2}|
\leq1/8$ they are suitably approximated by truncating at quadratic order in $\beta$ (at least for $\ell \leq 6$). We now make both statements quantitative.

How well a truncated power series in $\beta$ performs depends on the value of $\ell$. This manifests itself as a coupling between powers of $\beta$ and $\alpha_{2}$ for any given mass multipoles, which follows from counting polynomial degrees. That is, expanding the potential in Eq.~\eqref{eq:IntrinsicCurvature}, a term proportional to $\beta^{n}\alpha_{2}^{j}$ contributes a polynomial in $\zeta$ of degree at most $3n + 2j$, since the factor $1-\zeta^2$ adds two degrees and the two derivatives remove them. By the orthogonality of the Legendre polynomials, the general rule (which applies to any $\ell\geq2$) is that a term proportional to $\beta^{n}\alpha_{2}^{j}$ will contribute to $M_{\ell}$ only if
\begin{equation}\label{eq:IndexCondition}
    3n + 2j \geq \ell.
\end{equation}
The same argument applies to any Weyl distortion, as a term proportional to $\Pi_{n}\alpha_{n}^{k_{n}}$ contributes to $M_{\ell}$ only if $\sum_{n}nk_{n} \geq \ell$. Two consequences of this are worth highlighting.


The first concerns the relative size of the odd and even multipoles. As stated in Sec.~\ref{sec:level3.2}, the odd-parity component of any physically reasonable external source is likely to be much smaller than the reflection-symmetric component. What Eq.~\eqref{eq:IndexCondition} shows is that ``much smaller" must be measured against $\alpha_{2}^2$, not $\alpha_{2}$. The leading term of the odd multipole $I_{\ell+1}$ is proportional to $\beta\alpha_{2}^{(\ell-2)/2}$, while that of the even multipole $I_{\ell+2}$ is proportional to $\alpha_{2}^{(\ell+2)/2}$. Using Eqs.~\eqref{eq:DimensionlessMassMultipole},\eqref{eq:QuadDistortion-LeadingOrderMassMultipole} and \eqref{eq:BetaLinearOddMassMultipoles} at leading order one finds for even $\ell\geq2$
\begin{equation}
    \frac{I_{\ell+1}}{I_{\ell+2}} \approx C_{\ell}\frac{\beta}{\alpha_{2}^2},\quad C_{\ell} = \frac{\ell(\ell+2)\sqrt{(2\ell+3)(2\ell+5)}}{8(\ell+4)},
\end{equation}
so that $I_{\ell+1}$ overtakes $I_{\ell+2}$ once $|\beta| \gtrsim \alpha_{2}^2/C_{\ell}$, with $C_{2} \approx 1.3$, $C_{4} \approx 4.5$ and $C_{6} \approx 9.6$. $C_{\ell}$ grows as $\ell^2/4$ so the higher multipoles are the first to be dominated by the odd-parity component. Beyond $I_{2}$, one can therefore justify a pure quadrupole over a dipole-quadrupole-octupole distortion only if $|\beta| \ll \alpha_{2}^{2}/C_{\ell}$ for every multipole of interest, which is a far stronger condition than $|\beta| \ll |\alpha_{2}|$. The even multipoles are affected as well. At $|\alpha_{2}| = |\beta| = 1/8$ the $\beta$-independent term (the dashed curves in Fig.~\ref{fig:DPO_EvenLMoments}) is accurate to about 1\% for $M_{2}$, but misses $M_{4}$ by about 80\% and $M_{6}$ by order unity. The $\beta$-independent parts of $M_{4}$ and $M_{6}$ are suppressed by $\alpha_{2}^2$ and $\alpha_{2}^{3}$ respectively, while by Eq.~\eqref{eq:IndexCondition} their $\beta^2$ contributions are not suppressed by $\alpha_{2}$ at all.

\begin{figure}[htb]
    \centering
    \includegraphics[width=1.0\linewidth]{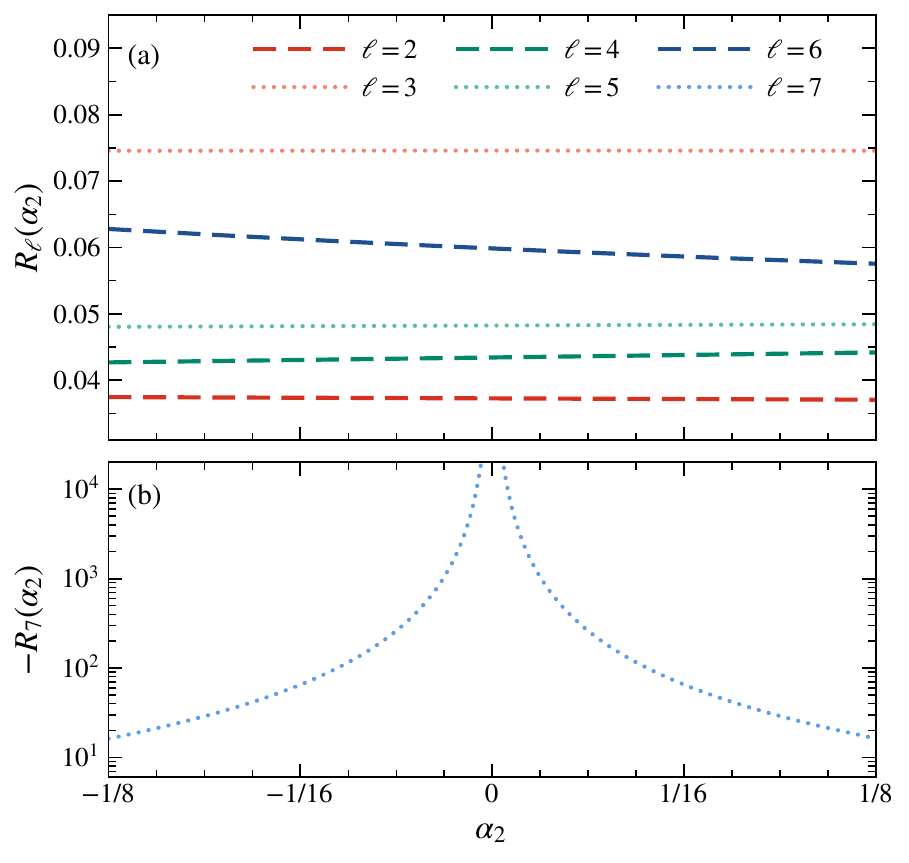}
    \caption{(a) A plot of $R_{\ell}(\alpha_{2})$ over $|\alpha_{2}| \leq 1/8$ for $\ell = 2-6$. The dashed lines represent even-$\ell$ modes while the dotted lines represent odd-$\ell$ modes. Each line is roughly constant and is determined by $\ell$. (b) An isolated plot of $-R_{7}(\alpha_{2})$ over $|\alpha_{2}| \leq 1/8$. As opposed to those curves in (a), $-R_{7}(\alpha_{2})$ diverges as $1/\alpha_{2}^{2}$. This is a direct consequence of Eq.~\eqref{eq:IndexCondition}.}
    \label{fig:R_Ell}
\end{figure}

The second consequence concerns the error of truncating the $n$-summation in Eq.~\eqref{eq:KEll-SeriesSolution}. Much like the leading-order analysis performed for the pure quadrupole distortion we can characterize this error by taking the ratio of coefficients of the leading order and first sub-leading order terms. For even $\ell$ this is the ratio of the $\beta^{4}$ and $\beta^{2}$ coefficients, while for odd $\ell$ it is the ratio of the $\beta^{3}$ and $\beta$ coefficients. For concreteness let us define the coefficient of the $\beta^{n}$ term in $M_{\ell}/M_{\neh}^{\ell+1}$ to be $c_{n}(\alpha_{2})$. Let us denote this ratio by $R_{\ell}(\alpha_{2}) := c_{n_0+2}/c_{n_0}$, with $n_0 = 2$ for even $\ell$ and $n_0 = 1$ for odd $\ell$. Let us further denote the truncated mass multipoles by $M_{\ell}^{(n_0)}$, then the relative error of the approximation is given by
\begin{equation}
    \frac{M_{\ell} - M_{\ell}^{(n_0)}}{M_{\ell}^{(n_0)} - M_{\ell}^{(Q)}} = \beta^2R_{\ell}(\alpha_{2}) + \mathcal{O}(\beta^4) \approx \beta^2R_{\ell}(\alpha_{2}).
\end{equation}
The above equation applies to both even and odd $\ell$ since $M_{\ell}^{(Q)}$ vanishes when $\ell$ is odd. It measures the error relative to the $\beta$-dependent part of the multipole, which is a stricter test than the error relative to $M_{\ell}$ itself. Figure \ref{fig:R_Ell} shows the value of $R_{\ell}(\alpha_{2})$ for $|\alpha_{2}| \leq 1/8$ for $\ell = 2-6$ (upper panel) and $\ell = 7$ (lower panel) so as to be compared to Figs.~\ref{fig:DPO_EvenLMoments} and \ref{fig:DPO_OddLMoments}. For $\ell \leq 6$, $R_{\ell}(\alpha_{2})$ is nearly independent of $\alpha_{2}$. Its size is set by $\ell$ and lies between 0.037 ($\ell=2$) and 0.075 ($\ell = 3$). The truncated multipoles shown in Figs.~\ref{fig:DPO_EvenLMoments} and \ref{fig:DPO_OddLMoments} for $\ell\leq 6$ are therefore accurate to $\beta^2R_{\ell} \leq 1.2\times10^{-3}$ over $|\beta|,|\alpha_{2}| \leq 1/8$. In particular, adding the $\beta^2$ term to the $\beta$-independent term brings $M_{2},M_{4}$ and $M_{6}$ at $|\alpha_{2}| = |\beta| = 1/8$ to within a relative error of $10^{-3}$.

This is not true at $\ell = 7$ however, as the lower panel of Fig.~\ref{fig:R_Ell} shows that $|R_{7}(\alpha_{2})|$ diverges as $1/\alpha_{2}^2$. This divergence is a direct consequence of Eq.~\eqref{eq:IndexCondition}. At leading order in $\alpha_{2}$ these coefficients scale as $c_{n}(\alpha_{2}) \propto \alpha_{2}^{j(n)}$ where $j(n) := \max(0,(\ell-3n)/2)$. For $\ell = 7$ the quantity $R_{7}(\alpha_{2})$ is determined by the ratio $c_{3}(\alpha_{2})/c_{1}(\alpha_{2})$, and if we take the leading order behavior we see that $|R_{7}(\alpha_{2})| \propto 1/\alpha_{2}^2$, since $c_{1} \propto \alpha_{2}^{2}$ while $c_{3} \propto \alpha_{2}^{0}$. Suppose we instead truncated at order $n = 3$ in $\beta$, then one finds that $R_{7}(\alpha_{2}) = c_{5}(\alpha_{2})/c_{3}(\alpha_{2}) \propto 1$ at leading order in $\alpha_{2}$. It becomes clear then that this coupling between $\beta$ and $\alpha_{2}$ can be an important source of error, but this can be reduced by including a few extra terms. The choice of truncation for the $n$-summation is therefore dependent on how many multipoles one wishes to accurately obtain, and Eq.~\eqref{eq:IndexCondition} determines which order in $\beta$ each multipole requires.

\section{Conclusion}\label{sec:level5}
In this work we have explored the horizon multipole moments of distorted Schwarzschild black holes. Starting from the exact integral of Ref.~\cite{Damour2009}, recast in the isolated-horizon framework developed by Ashtekar and collaborators \cite{Ashtekar2000,Ashtekar2002,Ashtekar2004}, we evaluate the geometric multipoles $I_{\ell}$ beyond linear order. For the case of a pure quadrupole distortion we are able to derive an exact analytical form for $I_{\ell}$ which simplifies to the linearized solutions found in the literature \cite{Damour2009}. We also consider a dipole-quadrupole-octupole distortion, for which we have not found an exact analytical form but a closed-form series solution is achievable. Such a solution expands $I_{\ell}$ in powers of the odd-parity distortion coefficient $\beta$, the coefficients of which are exact functions of the quadrupole distortion coefficient $\alpha_{2}$. Although the field multipoles of the distorted black hole are unchanged by the tides, these results show that its horizon multipoles are at least non-zero at every even $\ell$.

We then use these solutions (analytic, closed-form, and exact) to investigate the relative error of the approximate multipoles. In particular, we investigate the relative error of the leading-order approximations to the exact multipoles. For the pure quadrupole distortion model, we find that the leading-order approximation incurs a relative error that scales as $|\alpha_{2}|$ and is roughly independent of the $\ell$-mode when expressed in $\alpha_{2}$. In the case of linearization in $\alpha_{2}$, the only non-zero geometric multipole is $I_{2}$ and its relative error is given by $d_{2}\alpha_{2} = 6\alpha_{2}/7$. This error is larger than the first multipole that linearization neglects, so most of the error of pure linearization lies in $I_{2}$ itself, and the exact result of Eq.~\eqref{eq:J_Ell-ExactForm} removes it entirely.

When considering instead a dipole-quadrupole-octupole distortion, a term proportional to $\beta^{n}\alpha_{2}^{j}$ contributes to $M_{\ell}$ only if $3n+2j\geq\ell$ [Eq.~\eqref{eq:IndexCondition}]. Two consequences of this are noted. First, the odd-parity component must be small compared with $\alpha_{2}^2$, not $\alpha_{2}$, before it can be neglected. Second, the relative error for a truncation of the power series in $\beta$ goes as $\beta^2R_{\ell}(\alpha_{2})$, and $R_{\ell}(\alpha_{2})$ is nearly independent of $\alpha_{2}$ (see Fig.~\ref{fig:R_Ell}) so long as one truncates to a sufficient order in $\beta$ (as determined by Eq.~\eqref{eq:IndexCondition}). The value of $R_{\ell}(\alpha_{2})$ for $\ell \leq 6$ is at most 0.075, and so it follows that in the parameter space $|\alpha_{2}|,|\beta|\leq1/8$ the truncated approximations for the $\ell \leq 6$ multipoles are accurate to within order 0.1\% relative to their $\beta$-dependent parts.

The results derived in Sec.~\ref{sec:level3} can be used as benchmarks for horizon finders for any arbitrary distortions, including distortions well beyond linear theory. Since the leading-order error is roughly $\alpha_{2}$ at every $\ell$, linearied multipoles are reliable to a fractional accuracy $\varepsilon$ only for $|\alpha_{2}| \lesssim \varepsilon$. Eq.~\eqref{eq:GeometricMultipoleExactForm} provides an efficient way to calculate $I_{\ell}$ for any $\ell$ and to any precision, and Eq.~\eqref{eq:J_Ell-ExactForm} gives an algebraic form for $J_{\ell}(\alpha_{2})$ which provides a benchmark for those studying the pure quadrupole distortion, or who may be using it as a test case. However, this work is not without its limitations. The WdS spacetime is definitionally static and axisymmetric. While many astrophysical spacetimes can be justified as axisymmetric, the same is not always true for staticity. Many of the interesting cases studied are dynamical in nature \cite{Schnetter2006,PookKolb2020,Chen2022,Prasad2022} and in such systems the horizon evolves too. While the present work is not applicable to those systems, it is relevant to stationary spacetimes where the horizon remains an isolated horizon (for example, consider an astrophysical black hole with a rotating accretion disk with an inner boundary not on the horizon or a rotating distorted spacetime). In these cases the present work serves as analytical zeroth-order results, and in such a sense serve as benchmarks.

\acknowledgements
The author is greatful to Zhen Pan for providing comments on an early draft of this manuscript. This work makes use of Claude Opus 5.5 (Anthropic) to help in literature review, checking calculations, and suggesting revisions to the text. The author directed this work, verified all AI-assisted text edits and results, and takes full responsibility for the final form of the manuscript.

\bibliography{bibliography}
\end{document}